\documentclass[12pt]{iopart}

\usepackage{array}
\usepackage{graphicx}
\usepackage{cite}
\usepackage{url}
\usepackage{xspace}

\newcommand{\Pt}{\ensuremath{p_{\rm{T}}}\xspace}
\newcommand{\Et}{\ensuremath{E_{\rm{T}}}\xspace}
\newcommand{\Ht}{\ensuremath{H_{\rm{T}}}\xspace}
\newcommand{\Htcalo}{\ensuremath{H_{\rm{T}}^{\rm{calo}}}\xspace}
\newcommand{\Etmiss}{\ensuremath{E_{\rm{T}}^{\rm{miss}}}\xspace}
\newcommand{\Lumi}{\ensuremath{\mathcal{L}}\xspace}
\newcommand{\Orderof}[1]{\ensuremath{\mathcal{O}(#1)}\xspace}
\newlength{\tlen}

\begin{document}

\title{Triggering at High Luminosity Colliders}
\author{Hans Peter Beck}

\address{Laboratory of High Energy Physics, 
         University of Bern,
         Sidlerstr.~5,
         3012~Bern, CH}
\ead{hans.peter.beck@cern.ch}

\begin{abstract}
This article discusses the techniques used to select online promising 
events at high energy and high luminosity colliders.
After a brief introduction, explaining some general aspects of
triggering, the more specific implementation options for well
established machines like the Tevatron and Large Hadron Collider are
presented. An outlook on what difficulties need to be met is given when
designing trigger systems at the Super Large Hadron Collider, or at
the International Linear Collider.

\end{abstract}

\maketitle

\section{Introduction}

Exploring nature at higher and higher energies of particle collisions
has been proven to be a successful road to deepen our understanding of
the structure of matter. It has let to the formulation of the standard
model, which explains nature at its smallest scales and biggest
energy densities reachable by todays largest machines.
On the other hand, the standard model is almost certainly not the
final word of a theory describing nature at all scales and new
phenomena, not contained within the framework of the standard model,
necessarily exist. Such new phenomena are often referred to as \emph{new
physics}.

Colliding particles at the highest possible energies has the potential
to unveil manifestations of new physics, but these will necessarily be
covered within a huge background of already well known standard model
processes. Not only need the energy scale of the particle collisions
be pushed to the highest possible values, but also the collision rate,
i.e. the machine luminosity, needs to be pushed to unprecedented
values. In consequence, new or interesting physics events will be
covered by a large amount of well established standard model
processes. 

Todays largest collider complexes are the actual running Tevatron at
the Fermi National Accelerator Laboratory, Batavia, IL in the United
States and the soon to be completed LHC at CERN, Geneva, Switzerland.

Tevatron, which since its upgrade is referred to as Tevatron Run II
brings protons to collision with anti-protons at a centre of mass
energy of $\sqrt{s}\,=\,1.96$~TeV at a peak luminosity of
$\Lumi\,=\,2-3\,\times 10^{32}\,\rm{cm}^{-2}\rm{s}^{-1}$. The interval
between subsequent beam-crossings amounts to 396~ns, corresponding to
a beam-crossing rate of 2.5~MHz.

Two general-purpose experiments,
D\O~\cite{DO_Detector} and CDF~\cite{CDF_Detector} are recording those
proton--anti-proton collisions passing their respective trigger
schemes.

The Large Hadron Collider (LHC) is a proton-proton super-conducting
collider operating at $\sqrt{s}\,=\,14$~TeV at a nominal luminosity of
$\Lumi\,=\,10^{34}\,\rm{cm}^{-2}\rm{s}^{-1}$. The beam-crossing period
for proton-proton collisions at the LHC is 25 ns, corresponding to a
crossing rate of 40 MHz. At design-luminosity, $O(10^9)$ inelastic
proton-proton collisions will occur, implying an average of about 25
interactions per bunch crossing.

Two general-purpose
experiments, ATLAS~\cite{ATLAS_Detector} and CMS~\cite{CMS_Detector}
and three purpose built experiments, ALICE~\cite{ALICE_Detector},
LHCb~\cite{LHCb_Detector} and TOTEM~\cite{TOTEM_Detector}  are in
their final stages of installation and commissioning. LHCb is a
dedicated B-physics experiment with the aim of measuring the subtle
differences between matter and anti-matter. TOTEM is a dedicated
experiment to measure the total cross-section, elastic scattering and
diffractive processes of proton-proton collisions at the LHC.
ALICE will study the physical properties of matter during the early
stages after the big bang, for which LHC will also provide heavy-ion
lead-lead collisions.

\subsection{Physics goals, cross-sections and event rates}

At high luminosity hadron colliders discovery physics is the main
issue, which requires sensitivity of the trigger to a wide range of
signatures predicted by various theoretical models within and beyond
the Standard Model. 

\begin{figure}[htb]
  \centering
  \includegraphics[width=1.0\textwidth]{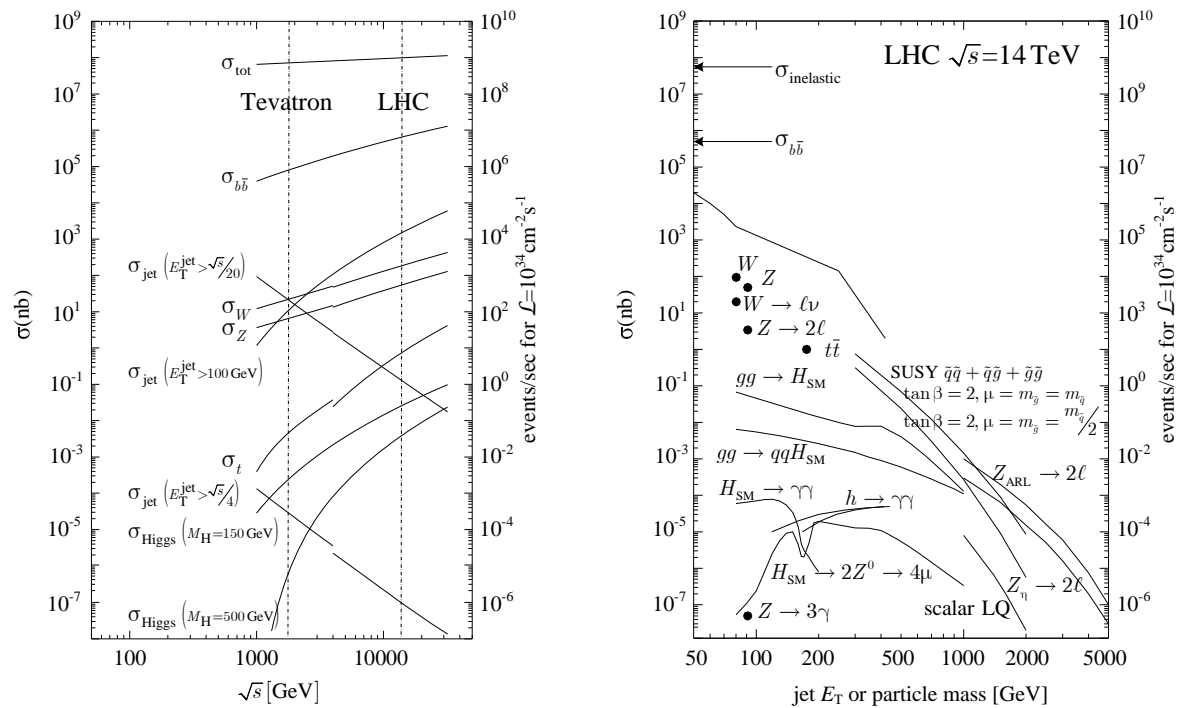}
\caption[Cross-sections and event rates] 
	{\label{fig:xsec}
	  Cross-sections and
	  event rates for proton proton collisions \\
	  left) as a function of the centre of mass energy 
	  covering Tevatron ($p-\bar{p}$ collision) and LHC ($p-p$
          collision) energy scales (figure reproduced from~\cite{ATLAS_TDAQ_TDR}).\\ 
	  right) as a function of the produced particle mass or the
	  highest jet transverse energy for $\sqrt{s}\,=\,14~\rm{TeV}$
	  (figure reproduced from~\cite{CMS_DAQ_PictureGallery}).\\ 
	}
\end{figure}

Cross-sections and event rates for proton proton collisions at the
Tevatron and LHC colliders are shown in figure~\ref{fig:xsec} as a
function of the produced particle mass or the highest jet transverse
energy. New physics, such as the yet to be discovered Higgs particle
or manifestations of super-symmetric particles happens in just one out
of \Orderof{10^8\cdots10^{13}} or even fewer proton-proton collisions.

Reading out the detector data at the fixed beam crossing rate and
analyzing the data off-line is beyond imagination of todays data
acquisition, data storage, and data processing systems. A powerful
selection mechanism is therefore needed to select, as early as
possible in the data acquisition chain, only those particle collisions
promising to contain interesting physics while simultaneously reducing
the overall data rate by many orders of magnitude.

This is the role of the trigger, which e.g. at the Tevatron detectors
CDF and D\O\ reduce the initial beam crossing rate by \Orderof{10^4},
and at the LHC detectors ATLAS and CMS by \Orderof{10^5}. One has to
note here that the number of proton-proton collisions is, due to
multiple interactions per beam crossing, in average $3-4$ times
larger than the beam crossing rate at the Tevatron collider and even
$\sim\!25$ times larger at the LHC.  Given that only one of these
multiple interactions give raise to a trigger signal, one can also
state that the actual physics selection of the trigger system selects
one proton-proton collision out of \Orderof{10^5} at CDF and D\O\ and
one out of \Orderof{10^7} at ATLAS and CMS. The data acquisition
system, however, can not distinguish between the actual proton-proton
collision that activated the trigger and the extra collisions that
took place within the same beam crossing. All proton-proton collisions
in a beam crossing are read-out simultaneously; in analogy of looking
at a photo that has been exposed multiple times. This
changes terminology of what is meant by an event. Often, an event is
referred to as all the tracks and energy deposits for a single
proton-proton collision. However, in high luminosity colliders, a
event is whatever tracks and energy deposits are read out per beam
crossing. Tracks and energy deposits, not originating from the primary
proton-proton collision that caused the trigger to fire, is called
\emph{pile-up}.

Precision measurements in contrast to discovery physics are often
severely disturbed in the presence of large pile-up. Therefore, data
taken when instantaneous luminosity is reduced; e.g. in the start-up
phase of a new accelerator (assuming an already well understood
detector) or at the end of an accelerator coasts, has much reduced
pile-up and thus provides a cleaner environment to allow for some
precision studies -- at the cost of a reduced statistical data set for
the analysis.

\section{Trigger strategies \label{TriggerStrategies}}

A powerful and flexible trigger is the cornerstone of a modern high
energy and high luminosity collider experiment. It dictates what
physics processes can be studied properly and what is ultimately left
unexplored.  The trigger must also offer sufficient flexibility to
respond to changing physics goals and to new ideas.  If the trigger is
not able to achieve sufficient selectivity to meet these requirements,
the physics potential of the experiment will be seriously compromised.

In an idealistic high energy physics experiment, enough bandwidth and
enough CPU power would be available to analyze every single
proton-proton collision and then decide whether the event contained
interesting physics.  In real life, available resources are limited
and trigger strategies based on the existence of one or few trigger
objects need to be implemented in a cost efficient manner.


\subsection{Trigger objects}

Features that distinguish interesting and/or new physics from the bulk
of Standard Model processes at colliders are typically the
presence of high-\Pt leptons $(e, \mu, \tau)$, photons, and jets, or
large missing transverse energy (neutrinos, or other weakly or non
interacting new particles), as they occur in very hard interactions or
as decay products from new heavy particles. Especially clean
signatures arise if the leptons, photons or non-interacting particles
are not covered inside jets but are \emph{isolated}.

\subsubsection{{\bf High-\Pt leptons}}
Triggering on high-\Pt leptons provides the primary means of selecting events
containing W, Z and Higgs bosons in $W\rightarrow l\nu$, $Z\rightarrow
l\bar{l}$, $H\rightarrow Z^{(*)}Z\rightarrow 4l$ decays, as well as
super-symmetric particles in $\chi_{2}^{0}\rightarrow l\tilde{l}$,
$\tilde{l}\rightarrow \chi_{1}^{0}l$ decays. Leptons can also tag
heavy quarks such as $t\rightarrow bW$ and $b$'s through their
semi-leptonic decays. In the latter, a $b$-tag can be obtained through
a lifetime selection of the relatively long lived B-meson, which leads
to non-pointing high-\Pt leptons with respect to the primary
interaction vertex.

Triggering on high-\Pt tau-leptons can lead to increase in statistical
samples, where the high-\Pt electron- and muon-trigger samples need to be
enhanced. However, the signature of tau's is all but clean due to
leptonic decays of the tau into a lighter lepton and two neutrinos
$\tau\rightarrow l\nu_l\nu_\tau$ or its semi-leptonic decays into hadrons
and a tau neutrino $\tau\rightarrow h\nu_\tau$.

In the pure leptonic case, the electron- or muon-trigger 
have a chance to select tau events in cases where the energy carried away
by the escaping neutrinos is not too high and the transverse momentum
\Pt of the decay lepton is still above the trigger threshold.  

Semi-leptonic tau-decays can be triggered by means of isolated,
high-\Pt tracks or very narrow jets. 

High-\Pt tau-triggers are more complex than other lepton triggers,
still they provide an important handle to new physics. For instance,
due to the relative strong coupling of tau to the Higgs particle,
which is enhanced with respect to the coupling to lighter leptons as
$m_\tau^2/m_l^2$, the decay mode $H \rightarrow \tau\tau$ is of
interest for Higgs discovery in the low Higgs mass
region~\cite{Higgs_tau_tau}.  In some of the large $\tan\beta$ SUSY
scenarios, $H\rightarrow \tau\tau$ is grossly enhanced over a wide
range of Higgs mass~\cite{MSSM_tau_tau}.

\subsubsection{{\bf High-\Pt photons}}
Triggering on high-\Pt photons is of prime interest for light Higgs searches,
such as $H\rightarrow\gamma\gamma$, as a very clean signal is
expected for Higgs masses below $\sim\!$120 GeV~\cite{ATLAS_Higgs_gg,
  CMS_Higgs_gg}.

On the other hand, high-\Pt photons occurring in QCD processes
via quark-gluon Compton scattering or quark-anti-quark annihilation
are often seen as background for the quest of new
physics. Nevertheless, these photons are useful in global fits
providing parton density functions to be used in Monte Carlo
simulations~\cite{ref_QCDphotons}.

Photons from highly energetic $\pi^0$ and $\eta$ mesons are a major
background to prompt photon signals. As these mesons are produced
within jets, an isolation criteria can be used to suppress such
photons. Photons originating from High-\Pt electrons undergoing
Bremsstrahlung are another source of background, which can be
suppressed after reconstruction of the electron track, and after
identification of a kink in the electron track.

\subsubsection{{\bf High-\Pt jets}}
Triggering on high-\Pt jets is hampered by the fact that jets are
produced abundantly in hadron colliders, as the standard QCD processes
$q+q\rightarrow q+q$, $q+g\rightarrow q+g$, $g+g\rightarrow g+g$ with
one or more extra gluons produced by QCD Bremsstrahlung.

High-\Pt jets which can hint for new physics occur from
$p+p\rightarrow X \rightarrow q+(q, l,...)$, with X being a new heavy
particle decaying hadronically into quark-quark, quark-lepton, or
quark + missing energy.

As the cross section for high-\Pt QCD jets is dropping more steeply
than the expected cross section for jets stemming from new heavy
particles they can be distinguished statistically in the high energy
end range of the jet energy spectrum.

It is therefore useful to look for combinations of multiple high-\Pt
jets, high-\Pt jets plus leptons or high-\Pt{}-jets plus missing
energy in order to reach conclusive results about possible heavy
states in proton-proton collisions.

Another possible way to improve the signal to noise ratio stemming
from QCD jets is to select those jets only that manifest from the
hadronization of a $b$-quark.  With the $b\bar{b}$ production cross
section being \Orderof{10^2} lower than the production of light
di-quarks, signal to noise ratios can be expected up to \Orderof{10}
enhanced, depending on the process under study.

Higgs searches in the low mass region relying solely on a $b$-tag to
search for $H \rightarrow bb$ events is close to impossible for
reasons of bandwidth and CPU power available for the trigger selection
processing, because the $b\bar{b}$ production rate is \Orderof{MHz}. 
In associated Higgs production, where besides the Higgs particle also two
top-quarks are produced, as in $p+p \rightarrow t+\bar{t}+H$, the $H
\rightarrow bb$ process can be searched for. This requires a
multi-object trigger, with asking for two $b$-tagged jets from the $H
\rightarrow bb$ decay, another two $b$-tagged jets from the two
top-quark decays with $t\rightarrow Wb$ and depending on the decay modes of
the W-boson a lepton tag from the $W\rightarrow l\nu$ decay or two
light-quark jets from the $W\rightarrow q_{u}q_{d}$ and $W\rightarrow
q_{c}q_{s}$ decays. 

\subsubsection{{\bf Missing energy}}
Triggering on missing energy is a window for new physics occurring from
$p+p\rightarrow X$, with $X$ being one or more new heavy particles
which are either stable and non-interacting with the surrounding
detector elements, or which decay into particles of which at least one
escapes detection due to non-interaction with the surrounding detector
elements. Important for missing energy is that the invisible particle
carries away a large amount of the available energy in transverse
direction to the beam line.  Missing energy combined with
leptons/photons or jets can be a manifestation of the presence of
large extra dimensions, different SUSY configurations, or other new
physics beyond the Standard Model.

Events containing multiple leptons and missing energy are often
referred to as the \emph{gold-plated} SUSY discovery mode. 

Obviously, missing energy can occur also due to detector
inefficiencies, dead areas, noise, as well as non-hermeticity of the
detector.
Still, triggering on missing energy is key for the quest in finding
new physics, which requires a detailed understanding of the detector
response.

\subsubsection{{\bf Total scalar sum of transverse energy}}
Triggering on total scalar sum of all transverse energy deposits in all
calorimeter cells and without further requirements on the event topology
allows for an open search for new heavy state particles.  However, the
summing up of all calorimetric cell energy deposits can be very
susceptible to both noise and pile-up effects and cannot be easily
calibrated unlike calibrating individual jet energies. Therefore a
modified total scalar transverse energy sum \Ht is used instead.
Experiments usually define \Htcalo as the
sum over the leading jet's transverse energies: 

\begin{equation*}
  \Htcalo = 
  \sum_{\rm{jets}}  E_{\rm{T}}^{\rm{jet}}
\end{equation*}

Often the leptons transverse energies and \Etmiss are also added:

\begin{equation*}
  \Ht = \Htcalo + \sum_{\rm{muons}} E_{\rm{T}}^{\mu} + \Etmiss
\end{equation*}

Note that electrons and to some extent taus, via the electron from the
$\tau \rightarrow e\nu_e\nu_\tau$ decay, are already accounted for in
\Htcalo.

The \Ht trigger can capture high jet multiplicity events
such as those from fully hadronic top decay, hadronic decays of
squarks and gluinos. These events have a total transverse energy of
several hundred GeV. They may actually fail the jet triggers because
individual jet transverse energies can be softer than the
sustainable thresholds of individual- and multi-jet triggers.

\subsection{Trigger chains}

In a first step, the trigger has to identify the trigger objects in
every event. Identifying e.g. an isolated high-\Pt electron and measuring at
least its transverse momentum is a multi-step procedure. In an initial
step, energy deposits in the trigger towers of the electro-magnetic
calorimeter need to be looked for and pattern recognition is needed to
identify close-by energy deposits as a single energy cluster and then to
determine the total energy deposit, and the centre of the cluster. Furthermore,
pattern recognition needs to identify whether the cluster is well
isolated from other energy deposits and especially whether it is well
isolated from energy deposits in the hadronic calorimeter, which
usually follows the electro-magnetic calorimeter.
After these steps, an isolated electromagnetic cluster is identified that could
either stem from an isolated electron or photon. Distinguishing the
two cases requires further analysis of the tracking system. In the
most trivial formulation, a cluster that stems from an electron has also
a corresponding electron-track pointing to it. However, as electrons
can undergo bremsstrahlung and as photons can convert into
electron-positron pairs, deciding whether an electro-magnetic cluster
stems from an electron is complex and resource intensive.

The trigger performs a chain of selection criteria on every trigger
object hypothesis to discriminate between promising
physics within the huge background. This is often called a
\emph{trigger chain} (sometimes also called \emph{trigger line} or
\emph{trigger path} by different experiments). For every trigger
object hypothesis one or more trigger chains are operated in parallel on every
event. This includes also combinations of trigger objects, where
e.g. two high-\Pt muons or jets are asked for.

\subsubsection{{\bf Exclusive trigger chains}}
select events according to some well known
properties of the wanted physics process, often including event
topology and invariant mass cuts, and usually lead to relatively small
accept rates. Trigger objects are combined at the trigger level
and an event is rejected if none of the wanted topologies is found. 

\subsubsection{{\bf Inclusive trigger chains}}
try to be as open as possible and select events based on the presence
of a single trigger object; e.g. a lepton with a transverse momentum
\Pt larger than a defined threshold.  Inclusive trigger chains can
lead to very high accept rates, even beyond the capacity of the data
acquisition system for detector read out. A remedy for this is to
raise the energy and momentum thresholds for these trigger objects
until acceptable rates are obtained that the data acquisition system
can handle. Needless to mention that doing so can lead to missing a
certain class of physics events.

Therefore, inclusive trigger chains at relatively low energy and
momentum thresholds need to be maintained. The high accept rates
are reduced based on a random selection; i.e. only every $N^{\rm{th}}$
event will be accepted for detector read out, which is often called
\emph{pre-scaling}.

\subsubsection{{\bf Trigger menus}}
are formed from all trigger chains, exclusive, inclusive, pre-scaled
and non-pre-scaled, that are operated together and in parallel during a
data taking run.

The trigger menu defines the strategy of selecting events, which are
believed in advance of being interesting. I.e. theoretical models and
detailed Monte Carlo simulations guide the definition of the trigger
menus. Nevertheless, the unexpected shall not be lost and thus trigger
menus usually contain a mixture of exclusive and inclusive trigger
chains. An example of a trigger menu for the CMS experiment is shown in
table~\ref{tab:CMSmenu} below~\cite{CMS_TrigTDR, CMS_HLT}:

\settowidth{\tlen}{.0}
\begin{table}
  \caption{\label{tab:CMSmenu}Example trigger menu for the CMS experiment
  for a luminosity of
  $\Lumi\,=\,2\,\times\,10^{32}\,\rm{cm}^{-2}\rm{s}^{-1}$. The L1
  rates shown are before applying pre-scaling.} 
  \begin{indented}
  \item[]\begin{tabular}{@{}l@{}rrrrr} \br 
    Trigger chain     & \multicolumn{1}{r}{L1 pre-scale}
                      & \multicolumn{1}{r}{L1 threshold}
                      & \multicolumn{1}{r}{L1 rate}
                      & \multicolumn{1}{r}{HLT threshold}
                      & \multicolumn{1}{r}{HLT rate} \cr
                      & 
                      & \multicolumn{1}{r}{[GeV]}
                      & \multicolumn{1}{r}{[kHz]}
                      & \multicolumn{1}{r}{[GeV]}
                      & \multicolumn{1}{r}{[Hz]} \cr
 \mr
    Inclusive electron &      1 &      22 &    4.2 &          29 &   24\hspace*{\tlen}   \cr
    Di-electron        &      1 &      11 &    1.1 &          12 &    1\hspace*{\tlen}   \cr
    Inclusive photon   &      1 &      22 &    4.2 &          80 &    3.1 \cr
    Di-photon          &      1 &      11 &    1.1 &      30, 20 &    1.6 \cr
    Inclusive muon     &      1 &      14 &    2.7 &          19 &   26\hspace*{\tlen}   \cr
    Di-muon            &      1 &       3 &    3.8 &           7 &    4.8 \cr
    Inclusive tau      &      1 &     100 &    1.9 &          -- &   --\hspace*{\tlen}  \cr
    Di-tau             &      1 &      66 &    1.8 &          -- &    6.0 \cr
    Single-jet         &      1 &     150 &    0.8 &         400 &    4.8 \cr
    Di-jet             &      1 &     100 &    1.7 &         350 &    3.9 \cr
    Triple-jet         &      1 &      70 &    0.7 &         195 &    1.1 \cr
    Quadruple-jet      &      1 &      50 &    0.5 &          80 &    8.9 \cr
    $b$-jet (leading jet) &   1 & 150, 100, 70, 50 & 1.8 & 350, 150, 55 & 10.3 \cr
    \Htcalo            &      1 &     300 &    1.2 &   -- &    --\hspace*{\tlen} \cr
    \Etmiss            &      1 &      60 &    0.4 &          91 &    2.5 \cr
    \Htcalo + \Etmiss  &      1 & 200, 40 &    0.7 &     350, 80 &    5.6 \cr
    tau + \Etmiss      &      1 &100 (tau)&    2.7 & 65 (\Etmiss)&    0.5 \cr
    jet + \Etmiss      &      1 & 100, 40 &    0.8 &     180, 80 &    3.2 \cr
    tau + electron     &      1 &  60, 15 &    2.6 &      52, 16 & $<$1.0 \cr
    tau + muon         &      1 &  40,  7 &    1.2 &      40, 15 & $<$1.0 \cr
    Inclusive photon   &    400 &      22 &    4.2 &          23 &    0.3 \cr
    Di-photon          &     20 &      11 &    1.1 &      12, 12 &    2.5 \cr
    Single-jet         &     10 &     140 &    1.1 &         250 &    5.2 \cr
    Single-jet         & $1\,000$   &      60 &   54\hspace*{\tlen}  &  120 &    1.6 \cr
    Single-jet         & $100\,000$ &      20 & 1718\hspace*{\tlen}  &   60 &    0.4 \cr
    \mr
    Total rate         &        &         &   23\hspace*{\tlen}  &      &  120\hspace*{\tlen} \cr       
    \br
  \end{tabular}
  \end{indented}
\end{table}

An important requirement for every trigger chain in any given trigger
menu is its \emph{selection efficiency}, which needs to be as high as
possible, bias free and known as precise as possible. Measuring
selection efficiencies from data is possible, where a trigger chain is
checked offline in a data sample where it was
not asked for the primary selection. Potential biases that can be
introduced by such methods need to be controlled and cross-checks with
pre-scaled trigger chains at reduced transverse energy and momentum
thresholds are needed. Often so called minimum bias events are used
for the determination of selection efficiencies. Since the
probability for finding a high-\Pt trigger object in a minimum-bias
data sample is usually very low, minimum-bias events are
only useful to measure selection efficiencies of trigger chains that
themselves require only a moderate-\Pt. Therefore, low- and/or
moderate-\Pt trigger objects need to be added to the trigger menu,
usually with a high pre-scaling. From there onwards, a recursive
procedure can be defined that allows to measure the selection
efficiency of all trigger chains.

The trigger menu must also ensure the allocation of adequate bandwidth
for calibration, monitoring, and background samples. These must be
provided in order to calibrate the detector and to control systematic errors.  

Balancing the event rates and the data volume, which can still be
handled by the data acquisition system, while maximizing the physics
reach of the experiment is just one of those optimizations that will
continuously need to be taken care of -- throughout
the lifetime of an experiment..

Trigger menus especially need to be adjusted following the accelerator
instantaneous luminosity, where (some of) the trigger chains need to
be pre-scaled or even be disabled completely according the actual
instantaneous luminosity, beam and detector conditions and taking into
account the priorities of physics goals defined by the collaboration. 

\section{Implementing a trigger and data acquisition system}

The trigger has to be capable of implementing the full trigger menu
with all trigger chains being executed in parallel.  Every trigger
chain in turn implies the execution of a number of sequential steps
that step by step validate the trigger chain or reject it. The initial
steps need to be executed at the bunch crossing rate dictated by the
accelerator, whereas later steps have more relaxed timing constraints.

It is therefore natural to implement a multi-level trigger to execute
the sequences of the trigger chains. The first level trigger (L1) has
to operate at the collision rate of the accelerator and usually cannot
be implemented using commodity components. An implementation based on
custom electronics and utilizing fast field programmable gate
arrays (FPGAs) and digital signal processors (DSPs) is unavoidable.
Only a small sub-sample of the detector data can realistically be fed
into the L1 trigger hardware, and only relatively simple algorithms
can be executed. Furthermore, changing algorithms at L1, other than
what can be done by re-configuration of pre-scales and threshold
values, is close to impossible or implies major upgrades.

For second and third level triggers (L2) and (L3) relaxed requirements
on decision latency and data volume exist. This allows the use of more
generic processing units and even personal computers (PCs) to be
used. The big advantage that follows from such an approach is the
utmost flexibility on the trigger algorithms that implement the
sequences of every trigger chain. New algorithms accounting for new
and better ideas to improve latency, efficiency and robustness for every
trigger chain can be added at any time throughout the experiment. Often,
triggers implemented based on PC farms are referred to as
\emph{high-level triggers} (HLT).

Experiments at Tevatron and LHC have chosen three trigger levels, with
the exception of the CMS experiment, which implements just two trigger
levels. Table~\ref{tab:TrigParams} shows the trigger parameters and
implementation choices of the CDF, D\O\ , ATLAS and CMS experiment.

\newcolumntype{R}[1]{>{\raggedleft}p{#1}}
\newcolumntype{C}[1]{>{\centering}p{#1}}
\newcolumntype{L}[1]{>{\raggedright}p{#1}}

\begin{table}
  \caption{\label{tab:TrigParams}Trigger parameters at Tevatron and LHC detectors.}
  \begin{indented}
  \setlength{\tlen}{.08\linewidth}
  \item[]\begin{tabular}{@{}l*4{@{}R{\tlen}@{\hspace{2pt}}L{\tlen}}} 
    \br 
                               & \multicolumn{4}{c}{Tevatron Run II}  & \multicolumn{4}{c}{LHC} \cr

    \mr
    pp centre of mass energy &
                                 \multicolumn{4}{c}{\makebox[0pt][r]{1.96\ }\makebox[0pt][l]{TeV}}   &  
                                 \multicolumn{4}{c}{\makebox[0pt][r]{14\ }\makebox[0pt][l]{TeV}}   \cr
    pp inelastic cross section &
                                 \multicolumn{4}{c}{\makebox[0pt][r]{50\ }\makebox[0pt][l]{mb}}   &  
                                 \multicolumn{4}{c}{\makebox[0pt][r]{70\ }\makebox[0pt][l]{mb}}   \cr
    Bunch crossing interval    &   
                                 \multicolumn{4}{c}{\makebox[0pt][r]{396\ }\makebox[0pt][l]{ns}}   &  
                                 \multicolumn{4}{c}{\makebox[0pt][r]{25\ }\makebox[0pt][l]{ns}}   \cr
    Bunch crossing rate        & 
                                 \multicolumn{4}{c}{\makebox[0pt][r]{2.5\ }\makebox[0pt][l]{MHz}}   &  
                                 \multicolumn{4}{c}{\makebox[0pt][r]{40\ }\makebox[0pt][l]{MHz}}   \cr
    Peak luminosity            &
                                 \multicolumn{4}{c}{\makebox[0pt][r]{$2-3\times10^{32}$\ }\makebox[0pt][l]{cm$^{-2}$s$^{-1}$}}  &  
                                 \multicolumn{4}{c}{\makebox[0pt][r]{$10^{34}$\ }\makebox[0pt][l]{cm$^{-2}$s$^{-1}$}}   \cr
    Number of bunches          & 
                                 \multicolumn{4}{c}{\makebox[0pt][r]{36\ }\makebox[0pt][l]{}}   &  
                                 \multicolumn{4}{c}{\makebox[0pt][r]{2808\ }\makebox[0pt][l]{}}   \cr
    Interactions per crossing  &
                                 \multicolumn{4}{c}{\makebox[0pt][r]{$3-4$\ }\makebox[0pt][l]{}}   &  
                                 \multicolumn{4}{c}{\makebox[0pt][r]{25\ }\makebox[0pt][l]{}}   \cr
    \br
    Detector                   & \multicolumn{2}{c}{CDF}   & \multicolumn{2}{c}{D\O\ }        
                               & \multicolumn{2}{c}{ATLAS} & \multicolumn{2}{c}{CMS} \cr
    Event size                 & 150 & kB      & 250 & kB     &  1.6 &  MB   & 1.0 & MB \cr
    \mr
    L1 signals                 &
                                 \multicolumn{4}{c}{calo/$\mu$/tracking}   & 
				 \multicolumn{4}{c}{calo/$\mu$} \cr
    L1 hardware                & \multicolumn{8}{c}{custom made electronics using ASICs, FPGAs and DSPs} \cr
    L1 rate                    &  10 & kHz    & 5 & kHz       & 100  & kHz     & 100 & kHz \cr
    L1 latency                 & 5.5 & $\mu$s & 4.2 & $\mu$s &  2.5  & $\mu$s &  3 & $\mu$s \cr
    \mr
    L2 signals                 & \multicolumn{4}{>{\centering}p{4\tlen}}{L1 information and extra detector information} &
                                 \multicolumn{2}{>{\centering}p{2\tlen}}{Region of Interests} &
				 -- & \cr
    L2 hardware                & \multicolumn{4}{>{\centering}p{4\tlen}}{custom electronics and generic processors} &
                                 \multicolumn{2}{c}{500 PCs} &
				 -- & \cr
    L2 rate                    &  350 & Hz     &   1 & kHz    &      3.5 & kHz & -- & \cr
    L2 latency                 & $\sim\!20$ & $\mu$s & $\sim\!100$ & $\mu$s & \Orderof{10} & ms & -- & \cr
    \mr
    L3 signals                 &  \multicolumn{8}{c}{full detector read-out fully digitized} \cr
    L3 hardware                &  100 & PCs    & 200 & PCs  & 1500 &    PCs & $\sim\!2000$ & PCs \cr
    L3 rate                    &  100 & Hz     &  50 & Hz   &  200 &  Hz & 120 & Hz \cr
    L3 latency                 & $\sim\!1$ & s  & $\sim\!1$ & s & \Orderof{1} & s & \Orderof{300} & ms \cr
    \br
  \end{tabular}
  \end{indented}
\end{table}

\subsection{L1 Trigger}
The L1 trigger has to deliver a new decision for every bunch-crossing,
which is much shorter time period than the latency it takes for L1
to operate.

A full latency determination starts at the instant the bunches
collide, and therefore includes the particle time-of-flight, the cable
propagation delay from detector to detector front-end electronics, and
the signal propagation time within the front-end electronics.  The
trigger must wait for the latest detector signal before processing can
begin, and in case a positive decision was taken, the LVL1 accept
signal needs to travel back to all detector elements to initiate
read-out. Considering a typical cable length from a sensitive detector
channel to the L1 trigger electronics and back to the detector of
\Orderof{200~\rm{m}} and assuming high quality cables capable of
transmitting electronic signals at a speed of~$\approx5~\rm{ns/m}$,
latency can not be kept below \Orderof{\rm{few}\,\mu\rm{s}}, see also
table~\ref{tab:TrigParams}.
 
\subsubsection{{\bf Selecting trigger objects at L1 trigger}}
The L1 trigger accepts candidate trigger objects, which are compatible
with the signature of high-\Pt leptons, photons and jets, as well as
missing energy and total scalar sum of transverse energy. All these
trigger objects deposit energy in the electro-magnetic and
hadronic calorimeters, with the exception of muons.
It is therefore sufficient to build a L1 trigger decision based on
information from the calorimetry and the muon system. 
However, the rate reduction achievable by L1 based on
calorimetric and muon system information alone may not be large enough, and
available bandwidth to the L2 trigger system and/or the available
computational resources at L2 can become a limiting factor.

The ATLAS and CMS experiments provide enough bandwidth and CPU
resources at their next trigger level, whereas the Tevatron
experiments D\O\ and CDF therefore implemented an L1 trigger system
that takes tracking into account.

D\O\ and CDF match muon track segments with tracks found in the inner
tracker  at L1 to reduced the L1 accept rate
further~\cite{CDF_L1track, D0_L1track}. 

This is detailed out in the following.

\subsubsection{{\bf L1 calorimeter trigger}}
Calorimeters, like e.g. the ATLAS calorimeter  have \Orderof{2 \times
10^5} cells to provide the granularity needed for proper event
reconstruction in a high-luminosity environment. Furthermore, raw
calorimeter signals extend over many beam crossings, which implies that
the information from a sequence of measurements of signal height in every
calorimeter cell needs to be combined in order to estimate the energy
deposit and to identify the beam crossing belonging to the  energy
deposit.

In order to reduce the data volume to be analyzed by the L1
calorimeter trigger, the analog signal of adjacent calorimeter cells
are summed to form $\sim\!7200$ \emph{trigger towers} at a typical
granularity of $\Delta\eta\times\Delta\varphi\,=\,0.1\times0.1$ in
pseudo-rapidity--azimuth space.

The trigger-tower signals are digitized using a dedicated ADC system
and digital signal processing is applied to extract the transverse
energy \Et for
calorimeter pulses and to assign it to the correct bunch crossing,
since the shaped pulses from the calorimeters extend over several
bunch-crossing periods

A sliding-window algorithm aims to find the optimum region of the
calorimeter for inclusion of energy from high-\Pt electrons, photons,
taus or isolated hadrons by moving a window grid across the
calorimeter space so as to maximize the transverse energy seen within
the window.
A second slightly different sliding window algorithm is performed to
find energy deposits originating from high-\Pt jets, which uses a
coarser granularity of
$\Delta\eta\times\Delta\varphi\,=\,0.2\times0.2$ in
pseudo-rapidity--azimuth space and different configurable 
window size. The optimum choice will depend on many factors: the jet
\Et of interest, the luminosity (level of pile-up within the window),
and the need to resolve nearby jets in multi-jet events.

These algorithms are performed in parallel at the beam-crossing rate of
40~MHz and with each trigger tower participation in the calculation of
up to 16 windows in both algorithms.

Summation is performed over the trigger towers to calculate the
\Etmiss vector and the total scalar \Et for the event. This is done by
summing the \Et values over all of the jet elements and the forward
calorimeters. In the case of the \Etmiss calculation, the vector
energy components are calculated from the \Et values, using lookup
tables to multiply by $\sin(\varphi)$ and $\cos(\varphi)$. After
summation of $E_x$ and $E_y$ separately, a look-up table is used to
compute the scalar \Etmiss value.

The total input bandwidth into the L1 calorimeter trigger system is
$\sim\!300$~GByte/s.

\subsubsection{{\bf L1 muon trigger}}
The L1 muon trigger has to identify with high efficiency genuine
high-\Pt muons, assign them to a particular beam crossing, and
determine their transverse momenta and location. 

\begin{figure}[htb]
  \centering
  \includegraphics[width=0.6\textwidth]{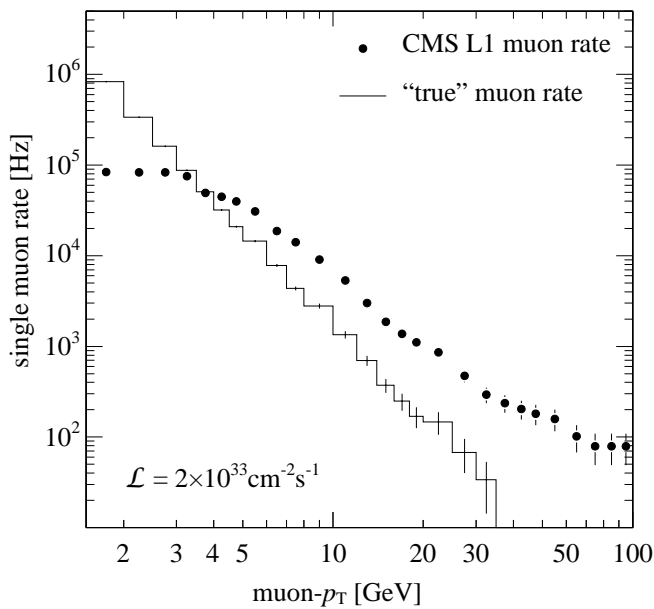}
  \caption[Cross-sections and event rates] {\label{fig:CMSL1muon} CMS L1
	trigger rate at
	$\Lumi\,=\,2\,\times\,10^{33}\,\rm{cm}^{-2}\rm{s}^{-1}$ as a
	function of \Pt threshold for single-muons (figure reproduced
	from~\cite{CMS_DAQ_PictureGallery}). 
       }
\end{figure}

The rate of muons produced in LHC collisions at design luminosity is
enormous \Orderof{1}~MHz for small muon-\Pt, as is shown in
figure~\ref{fig:CMSL1muon}.

Muon systems, like e.g. the one for CMS comprise multiple sub-systems
with the choice of the detector technologies being driven by the very
large surface to be covered and by the different radiation environments.

In the CMS barrel region ($|\eta| < 1.2$), where the neutron
induced background is small, the muon rate is low and the residual
magnetic field in the chambers is low, drift tube (DT) chambers are
used. In the two end-caps, where the muon rate as well as the neutron
induced background rate is high, and the magnetic field is also high,
cathode strip chambers (CSC) are deployed and cover the region up to
$|\eta| < 2.4$. In addition to this, resistive plate chambers (RPC) are
used in both the barrel and the end-cap.

The L1 muon trigger of the CMS experiment is based on custom
electronics and takes signals from all three muon chamber systems.

The L1 electronics for the DT and CSS chambers first process the
information locally, delivering position, direction, bunch
crossing, and quality per muon candidate object. Such candidate
objects produced by local triggers are often referred to as \emph{trigger
primitives}.

The L1 electronics for the RPCs is based on the spatial and time
coincidence of hits in four RPC muon stations. The candidate track is
formed by a pattern of hits that matches with one of many possible
patterns pre-defined for muons with defined transverse momenta. The \Pt
value is thus given. 

Trigger primitives from the DT, CSC and RPC L1 electronics are
collected and converted into muon tracks where also transverse
momentum \Pt, pseudo-rapidity $\eta$ and azimuth $\varphi$ are
assigned. In addition, the $\eta-\varphi$ coordinates are correlated
with the signals from the calorimetric towers to decide whether the
muons are isolated. 

The CMS L1 trigger electronic is limited to a maximum accept rate of
$10^5$~Hz and delivers trigger decisions in 3~$\mu$s at the
beam-crossing rate of 40~MHz. Random coincidences, energy-loss
fluctuations and multiple-scattering need to be taken into account in
the L1 trigger logic and softens the sharpness with which a muon's \Pt
can be determined. The effect of this can also be seen in
figure~\ref{fig:CMSL1muon}, where the CMS L1 muon rate is shown as a
function of the muon-\Pt. The L1 muon rate is constant just below
$10^5$~Hz for muon-\Pt$<3.5$~GeV and falls off softer than the ``true''
muon rate, which has been generated using Monte-Carlo simulation, for
larger values of \Pt. Even at a \Pt threshold of 100~GeV, 
where signal muon events will be rare, close to 100~Hz of muon L1
trigger accept rate is expected.
A higher level trigger decision is thus needed to re-analyze these muons
with greater precision.

\subsubsection{{\bf L1 tracking trigger}}
The ATLAS and CMS experiments at the LHC did not chose to build a L1
tracking trigger. However, the CDF and D\O\ experiments at Tevatron
does trigger on tracks already at L1.

The reason for ATLAS and CMS of not implementing tracking triggers at
L1 are primarily due to involved complexity and extra costs for with
respect to the expected gain in physics coverage. With an L1 accept
rate of up to 100~kHz and a data acquisition system providing ample
bandwidth to move detector data following a L1 accept to the next
trigger level, extra tracking information at L1 becomes redundant.
Only in the case where low-\Pt physics, where all particles produced
in a collision are below threshold of $\Pt\approx3-6$~GeV, would need
to be studied more carefully, tracking at L1 becomes relevant.
The Standard Model use case motivating tracking rigger at L1 is
B-physics, where a b-quark fragments into a relatively long-lived
B-meson that decays either hadronically or leptonically into particles
showing a common displaced vertex with respect to the main $p-p$
collision vertex. Looking again at figure~\ref{fig:xsec}, one recognizes
a total rate of $b\bar{b}$ production of close to $\sim\!10^6$~Hz, which
would be unimaginable to handle in a general purpose experiment as
ATLAS or CMS. The LHCb experiment is specially built to measure
B-physics and its first trigger level has indeed an accept rate of
$10^6$~Hz~\cite{LHCb_TDAQ_TDR}.
ATLAS and CMS will play their role in B-physics whenever the
B-particle decays in a sufficiently high-\Pt muon or
muon-pair~\cite{Atlas_Bphys, CMS_Bphys}. 

At Tevatron the CDF and D\O\ experiments both implement tracking
triggers at L1~\cite{CDF_L1track, D0_L1track}, which allows them to
further reduce the L1 accept rate. Fake muon tracks selected by the L1
muon system, but due to random coincidences, neutron background or
from real muons that stem from $K^0_L\rightarrow\pi\mu\nu_\mu$ decays
will thus no-longer contribute to the overall L1 accept rate.  

In addition, the CDF experiment also identifies displaced vertices
already at L1 which improves the B-physics potential of the CDF
experiment.

\subsubsection{{\bf L1 global trigger}}
Trigger objects identified by the L1 calorimeter, muon and, if
applicable, tracking trigger need to be combined and matched against
the trigger menu before a global L1 trigger decision can be
taken. This is the task of the \emph{L1 global trigger} (sometimes
also called \emph{L1 central trigger} by different experiments). In
most experiments this is done by counting the number of isolated and
non-isolated L1 trigger objects that are above a predefined set of
transverse energy and transverse momentum threshold. In most
experiments no exclusive trigger chains can thus be applied in L1
trigger menus.

A special feature worth noting is the capability of the CMS L1 global
trigger which also considers topological features of trigger objects
based on pseudo-rapidity $\eta$ and azimuth $\varphi$
information~\cite{CMS_L1Global}. The CMS L1 trigger menu can thus be
enriched with a set of exclusive trigger chains. An example being e.g. two
isolated electrons or muons with a back-to-back signature and opposite
charge, or any other relative opening angle. Although the ATLAS L1
trigger provides extra geometry information with every L1 trigger
object, this information, however available, is not further considered
at L1 but directly passed to the L2 trigger.

\subsection{L2 Trigger}
The strategy of L2 triggers is to refine L1 trigger decisions and to
combine trigger objects identified by L1, e.g. in combining data across
detectors to form higher quality trigger objects and examining
event-wide correlations in all L2 trigger objects.

The D\O\ and CDF experiments pass  the L1 trigger objects and
some detector data to the L2 electronics. The L2 triggers for D\O\ and
CDF are both based on a mixture of custom built electronic boards and
general purpose processors accepting input signals at the respective
L1 accept rates of $\sim\!5$~kHz in the case of D\O\ and $\sim\!10$~kHz in
the case of CDF. A L2 decision is taken in $\sim\!100\mu$s to reduce the
event rate further down to $\sim\!1$~kHz at D\O\ and in $\sim\!20\mu$s to
reduce the event rate further down to $\sim\!350$~kHz at CDF.

The differences in the L2 accept rates and latencies for the two
experiments is a mere choice of the collaboration.
Large latencies imply more time to execute the selection steps for
every trigger chain of the trigger menu on the cost of providing
larger buffering and also larger processing environment in proportion
to the L1 accept rate times latency. 

When a large latency of \Orderof{10}~ms or higher can be afforded, an
interesting possibility opens. It becomes possible to implement the
execution of the selection steps in a pure software based environment,
providing utmost flexibility in adopting selection strategies and
improving trigger menus according the evolving needs and requirements
that come up during the whole running time of the experiment.

This has been the choice of the ATLAS and CMS experiment.

The ATLAS L2 trigger is a farm of \Orderof{500}~PCs equipped with dual
multi-core CPUs. As PCs are commodity devices, they can be easily replaced
with faster models and thus a continuous improvement of the total
available processing power can be expected.

Every L2 PC in ATLAS treats one or more L1 accepted events in
parallel. For every event, information from the L1 trigger is passed
to a L2 PC, which consists of so called Region of Interest (RoI)
information. The RoI information contains meta data of the L1 trigger
objects found, such as trigger object types e.g. electron/photon clusters,
jets, \Etmiss, muons and their respective trigger thresholds passed,
and for every trigger object also their coordinates in pseudo-rapidity
$\eta$ and azimuth $\varphi$ space. The L2 PC subsequently collects
detector data from the ATLAS read-out system in order to confirm or
reject one L1 trigger object after the other, and to refine trigger
objects with extra detector data; e.g. in finding tracks from
the inner tracker that match with a cluster found in the calorimeter
or with a muon track found in the muon system.
An average latency of $\sim\!10\,\mu$s of CPU time spent per event is
deemed to be sufficient to execute the ATLAS L2 selection. With most
events being rejected within $\sim\!2-3\,\mu$s, and a correspondingly
larger time budget available for more interesting events.

Only about 2\% of the data volume of 1.6~MB per event needs to be
moved at L1 rate from the detector read-out system to the L2 trigger
and analyzed, corresponding to \Orderof{3}~GB/s. A dedicated commodity
gigabit Ethernet network can handle such data volumes. 

An important difference between the ATLAS and the CMS experiment is
that the CMS experiment decided against a dedicated L2 trigger as CMS performs
the next selection step directly at L3.

\subsection{L3 trigger}

In all experiments at Tevatron and LHC, the L3 trigger is a computer
farm that uses software algorithms for particle identification after
event reconstruction. The final rate for writing events to tape is
between $50-200$~Hz for the various experiments.

High level triggers (HLTs) are PC farms that are flexible, that do not
require special infrastructure for the development of trigger
algorithms other than a laptop. In principle complex offline-like 
algorithms can be executed. On the other side, many hundreds or
thousands of PCs need to be operated reliably in a huge
farm. Assuring a proper load-balancing when assigning an event to
be executed on an individual PC requires non-trivial management of the
data flow that also needs to be robust against non-responding PCs due
to crashes or other failures. 

The CMS experiment decided for a HLT farm capable of absorbing the
events at the 100~kHz accept rate of their L1 trigger. The main problem
for this is again the data movement, as at a L1 event rate of 100~kHz
and a typical event size of 1~MB a total data volume of 100~GB/s
arises. The final size of the farm is not yet determined as this
depends a lot on the availability of fast multi-core CPUs and the
execution speed of the selection software. At an average latency of
$\sim\!300$~ms per event, one estimates $30\,000$~CPU cores that need to
be operated in parallel~\cite{SergioCittolin}. 
Today, PCs with 8~CPU cores are easily available and PCs with 16 and
more CPU cores are expected within a year. This translates to
\Orderof{2000}~PCs to be operated.

Also ATLAS plans to deploy 2000~PCs for its HLT operation, with
$\sim\!500$~PCs being used for the L2 trigger and $\sim\!1500$~PCs
utilized for the L3 selection.

At the time of this write-up it is not yet clear which of the two
approaches will result as the better choice for the experiments
trigger selection. As the trigger chains require sequential steps to
be executed, the decision taking processes become equivalent in both
experiments. The difference however is that in the case of CMS the
complete event data is available for processing already after L1,
which in principle allows to play sophisticated tricks with some
exclusive selection of lower-\Pt physics events that otherwise would not
have been selected online -- assuming availability of enough CPU
resources to execute more trigger chains in parallel.
Note that also the ATLAS L2 trigger has access to all detector data -
in principle. With more CPU resources available and an upgrade of the
commodity gigabit Ethernet network, low-\Pt track searches could also be
performed in ATLAS.

\subsection{Triggering the unexpected}
A question that is often asked is whether it is possible to discover
physics that was not thought for when defining the trigger chains and
when setting up the trigger menu.
There is no simple answer to this question. As can be seen from the
trigger menu presented in table~\ref{tab:CMSmenu}, simple trigger
chains have been defined for selecting single-jet events at L1 with a
transverse momentum as low as \Pt=\,20~GeV. However, a huge pre-scale
value of $100\,000$ is required to yield an accept rate of 1.7~kHz. If
there are enough CPU resources available at the following selection
steps, a detailed search for low-\Pt phenomena can still be done --
with a corresponding loss of available luminosity for such a trigger
chain, as given by the pre-scale value.

Another exotic case that can be difficult to trigger on are the
production of slow but heavy particles that only decay after having
traveled many centimeters or even meters into the detector~\cite{SMP}.
As these particles are supposedly slow, they may or may not be
associated in the read-out with their original beam-crossing and could
appear as a stand-alone events. The original event may or may not show
enough missing transverse energy \Et and the decay products of this
heavy particle are likely not to match any of the trigger masks that
would execute a L1 accept, especially if it decays into muons or even
worse, into neutrinos. Thus there is always room for the
unexpected to escape detection, even in the most sophisticated trigger
systems of today.

\section{Conclusion and outlook }

A powerful and flexible trigger system is decisive in a modern high
luminosity collider experiment. It dictates on the physics
processes that can be explored and on what is ultimately left
unexplored.

Trigger menus need to be carefully composed from trigger chains, which
ultimately identify the trigger objects in a sequence of processing
steps that define the selection process and span over multiple trigger
levels. Discarding unwanted trigger objects early in a sequence of
selection steps is key in reducing the overall requirements for
bandwidth and CPU resources. The largest fraction of rate reduction is
therefore usually performed at the L1 trigger level, which, due to its
hardware-based implementation, and fixed wiring, also provides the
most rigid infrastructure for event selection.

With the availability of higher bandwidth for read-out and the massive
amount of CPU resources that can be provided by means of large CPU
farms, the event selection is, wherever possible, no longer
performed in a custom built electronics environment, but in a flexible
and adaptive software environment, executing selection processing on
many hundred or thousands of PCs, all working in parallel.

This trend will continue, as it provides the most flexible environment
to adopt new trigger strategies, and to compose new trigger menus
containing new ideas for individual trigger chains; e.g. taking into
account complex event topologies.

\subsubsection{{\bf Super LHC}}
The LHC accelerator is proposed to be upgraded to the Super-LHC
(SLHC)~\cite{SLHC}, where luminosity will be increased by a factor of
ten to \Lumi~=~$10^{35}$~cm$^{-2}$s$^{-1}$ while the proton-proton centre
of mass energy will remain at $\sqrt{s}~=~14$~TeV. The beam-crossing
interval is likely to double to 50~ns, with its final value not yet
concluded. Assuming 50~ns, an average of 500~pile-up events will occur
with every beam-crossing, which will lead to higher occupancy and
radiation-levels in the detector systems. Replacing some of the
detector infrastructure, notably the tracking devices, will likely be
required.  Regarding the trigger systems, the L1 calorimetric triggers
of ATLAS and CMS may just need to be raise their thresholds, whereas the
L1 muon triggers will suffer from a higher fake muon rate.  This is
especially true for the CMS muon system, where multiple scattering of
muons in the iron-core has lead to the definition of many more
patterns that need to be matched against hits found in the CMS muon
systems to form L1 muon tracks. At SLHC, the L1 muon fake rate is
likely to be a major issue and a match with L1 tracks from a new to be
built inner tracker is being considered~\cite{CMS_L1track}.

The ATLAS L1 muon system is much less exposed to multiple scattering
effects due to its air-core toroid magnet. Therefore, the ATLAS L1
muon trigger is thought to be able to cope even with a ten-fold
increase of luminosity.

Average event size will be higher at SLHC, due to higher occupancy in
the detector elements. The read-out elements and bandwidth to the PC
farms for the higher level triggers, as well as their CPU resources
will need to be upgraded. However, the overall strategy, as it has
been described in this paper, provides enough flexibility to be also
valid at SLHC. 

\subsubsection{{\bf International Linear Collider}}

The International Linear Collider (ILC) global design~\cite{GDE}
effort proposes 
the next high energy and high luminosity collider to be an
electron-positron linear collider at a centre of mass energy of
$500-1000$~GeV at a luminosity of
$\Lumi\,=\,2\,\times\,10^{34}$~cm$^{-2}$s$^{-1}$. The beam will arrive
in a train of 2625 bunches within 970~$\mu$s, with a bunch spacing of
370~ns. Every 200~ms a new train of 2625 bunches will arrive at the
collision point, which leaves an interval of 199.3~ms with no collisions.
For the trigger and data-acquisition system at a future ILC
experiment, an interesting possibility opens. During the 970~$\mu$s, where
collisions take place every 370~ns, all the detector signals can be
kept in on-detector memories. During the 199.3~ms with no collisions,
all the detector signals can be read out and sent to a big PC farm,
where the complete trigger menu can be executed. A
hardware based L1 trigger, with its rigidity against adopting trigger
strategies, is no longer needed; assuming that enough bandwidth for
the read-out and enough CPU resources for the execution of the trigger
menu can be provided. 

As pointed out in section~\ref{TriggerStrategies}, the \emph{ideal}
trigger for a high energy and high luminosity collider can become
reality at future ILC experiments. 

\ack
The author likes thank his many colleagues widely spread over many
experiments from D\O , CDF, ATLAS and CMS. Without having tea
or coffee with many of them at the CERN cafeteria and elsewhere, this
article would not have been possible to realize.

\end{document}